\documentclass[twocolumn]{aastex7}
\usepackage{kotex}
\usepackage{multirow}
\usepackage{gensymb}
\usepackage{amsmath}
\usepackage{graphicx}
\usepackage{xcolor}
\usepackage[T1]{fontenc} 
\usepackage{enumitem}
\usepackage{comment}

\begin{document}

\title{XRISM/Resolve Spectroscopy of the Central Engine in the Seyfert-1 AGN Mrk 279}

\author[orcid=0000-0003-2869-7682]{Jon M. Miller}
\affiliation{Department of Astronomy, University of Michigan, MI 48109, USA}
\email[show]{jonmm@umich.edu}

\author[0000-0002-7129-4654]{Xin Xiang}
\affiliation{Department of Astronomy, University of Michigan, MI 48109, USA}
\email{xinxiang@umich.edu}

\author[orcid=0000-0002-3687-6552]{Doyee Byun (변도의)}
\affiliation{Department of Astronomy, University of Michigan, MI 48109, USA}
\email{doyeeb@umich.edu}

\author[0000-0001-9735-4873]{Ehud Behar}
\affiliation{Department of Physics, Technion, Technion City, Haifa 3200003, Israel}
\email{behar@physics.technion.ac.il}

\author[0000-0003-2663-1954]{Laura Brenneman}
\affiliation{Center for Astrophysics | Harvard-Smithsonian, MA 02138, USA}
\email{lbrenneman@cfa.harvard.edu}

\author[0000-0002-8294-9281]{Edward Cackett}
\affiliation{Department of Physics and Astronomy, Wayne State University, Detroit, MI 48201, USA}
\email{ecackett@wayne.edu}

\author[0000-0001-8470-749X]{Elisa Costantini}
\affiliation{SRON Space Research Organization Netherlands, Niels Bohrweg 4, 2333CA Leiden, The Netherlands} 
\affiliation{Anton Pannekoek Institute for Astronomy, University of Amsterdam, Science Park 904, NL-1098 XH Amsterdam, The Netherlands}
\email{E.Costantini@sron.nl}

\author[0009-0006-4968-7108]{Luigi Gallo}
\affiliation{Department of Astronomy and Physics, Saint Mary's University, Nova Scotia B3H 3C3, Canada}
\email{lgallo@ap.smu.ca}

\author[0000-0003-1728-0304]{Keith Horne}
\affiliation{SUPA Physics and Astronomy, University of St. Andrews, Fife, KY16 9SS, UK}
\email{kdh1@st-andrews.ac.uk}

\author[0000-0002-0273-218X]{Elias Kammoun}
\affiliation{Cahill Center for Astrophysics, California Institute of Technology, Pasadena, CA 91125, USA}
\email{ekammoun.astro@gmail.com}

\author[0000-0001-5493-7585]{Chen Li}
\affiliation{Leiden Observatory, Leiden University, PO Box 9513, 2300 RA Leiden, The Netherlands}
\email{cli@strw.leidenuniv.nl}

\author[0000-0002-0572-9613]{Abderahmen Zoghbi}
\affiliation{Department of Astronomy, The University of Maryland, College Park, MD 20742, USA}
\affiliation{HEASARC, Code 6601, NASA/GSFC, Greenbelt, MD 20771, USA}
\affiliation{CRESST II, NASA Goddard Space Flight Center, Greenbelt, MD 20771, USA}
\email{azoghbi@umd.edu}

\begin{abstract}
High-resolution X-ray spectroscopy with XRISM gives an unprecedented
view of the ``central engine'' in active galactic nuclei, providing
unique insights into black hole accretion and feedback.  We present an
analysis of the first XRISM/Resolve spectrum of the Seyfert-1 galaxy
Mrk~279, known for its complex line profiles and variability.  The
data reveal velocity components within the Fe~K$_{\alpha}$ emission
line that can be associated with the inner face of the molecular torus
($r \geq 10^{4}~GM/c^{2})$, the broad line region (BLR; $r =
1650^{+5780}_{-1480}~GM/c^{2}$), and the inner accretion disk ($r =
81^{+280}_{-75}~GM/c^{2}$).  We find evidence of low-velocity, highly
ionized gas that contributes an H-like Fe~XXVI emission line at
6.97~keV, confirming suggestions from prior low-resolution spectra.
The data do not show slow winds in absorption, but two
  pairs of lines -- consistent with He-like and H-like Fe
  shifted by $v\simeq 0.22c$ and $v\simeq 0.33c$ -- improve the fit,
  and could represent an ultra-fast outflow (UFO).  Their addition to
  the model only reduces the Akaike Information Criterion by 3.6 and
  3.5, respectively, signaling modest support. Additional
observations are needed to definitively test for the presence of fast
X-ray winds in Mrk~279. We discuss these results in the context of the
geometry of the central engine in AGN, emerging trends in XRISM
studies of AGN, and the nature of the potential UFOs.
\end{abstract}


\section{Introduction}

The region between the innermost stable circular orbit (ISCO) and the
molecular torus is sometimes referred to as the ``central engine'' in
active galactic nuclei (AGN).  It is within this expanse -- perhaps
$\Delta r \simeq 10^{5}~GM/c^{2}$ -- that accretion generates
radiative and mechanical feedback that can reshape host galaxies.
Ironically, although interferometry has imaged the shadow of the event
horizon and the torus (e.g., \citealt{eht2019},\citealt{gravity2024}
), we are not yet able to directly image the orders of magnitude in
between these extremes.  Spectroscopy, variability, and other tools
must be harnessed to understand this region.

In some key respects, X-rays may be the most incisive probe of the
central engine.  Unlike the UV and EUV light that dominate the
radiative output from AGN, X-rays are not as easily attenuated by gas
and dust, X-ray lines are excited by the X-ray continuum captured by
the same instrument, and the host galaxy does not contribute much
background.  With the launch of XRISM, X-ray spectroscopy has taken a
major step forward.  The Resolve calorimeter delivers a resolution of
$\Delta E = 4.5$~eV in the Fe~K band, or $R = E/\Delta E \geq 1400$
(for descriptions of XRISM and its instruments, see
\citealt{tashiro2021}, \citealt{ishisaki2022},
\citealt{hayashida2018}).  It is in this band that the neutral
Fe~K$_{\alpha}$ line is found -- the best X-ray diagnostic of the
distribution of cold gas -- as well as the most powerful, highly
ionized winds (for a review, see \citealt{gallo2023}).

XRISM spectroscopy of the nearby Seyfert-1.5 galaxy NGC~4151 reveals separate contributions to the ~Fe~K$_{\alpha}$ line from radii that are broadly consistent with the inner face of the molecular torus, the optical ``broad line region'' (or, BLR), and the disk close to the ISCO \citep{2024ApJ...973L..25X}.  The same observations reveal a rich wind spectrum, with component velocities ranging from $2 \times 10^{2} \leq v \leq 5\times 10^{4}~{\rm km}~{\rm s}^{-1}$ \citep{xiang2025}.  The kinetic power of the fastest wind components appears to exceed the threshold at 0.5-5\% of the Eddington limit, identified in simulations as the threshold for reshaping host galaxy structure (\citealt{dimatteo2005}, \citealt{hopkins2010}).

It is possible that Seyfert-1.5 AGN afford a fortuitous view of the central engine, sampling a large range of radii, geometries, and physical processes, but additional observations of related AGN are needed to confirm this.  To this end, we observed the Seyfert-1 galaxy Mrk~279 ($z=0.0305$, \citealt{strauss1988}) during XRISM Cycle 1.  At $M_{\rm BH} = 3.8^{+0.9}_{-0.6}\times 10^{7}~M_{\odot}$ \citep{williams2018}, the mass of the black hole in Mrk~279 is similar to that in NGC 4151 ($M_{\rm BH} = 3.4\pm0.4\times 10^{7}~M_{\odot}$; \citealt{bentz2015}), but its Eddington fraction is potentially an order of magnitude higher \citep{vasudevan2007}.  Importantly, prior observations with XMM-Newton suggest a composite Fe~K$_{\alpha}$ emission line in Mrk~279 \citep{costantini2010}, and Chandra spectra reveal broadened He-like O~VII lines that likely originate in the BLR \citep{costantini2007}.  

In this paper, we report on the first XRISM observation of Mrk~279.  Section 2 describes the key observation details and how the data were reduced.  Our analysis procedure and results are described in Section 3.  Finally, in Section 4, we discuss our results and their impact on our understanding of Mrk~279 specifically and AGN central engines in general.

\section{Observation and Data Reduction}
\label{sec:observation}
XRISM observed Mrk~279 starting on 2024 September 19, starting at 19:58:04 UTC.  The high-resolution Resolve calorimeter \citep{ishisaki2022} was run in its standard ``PX\_NORMAL'' mode.  The Xtend CCD camera \citep{hayashida2018} was run in the ``Window1'' mode to limit photon pile-up in the data.  Prior studies with XMM-Newton have observed Mrk~279 at CCD resolution (e.g., \citealt{costantini2010}), so this work is focused on the Resolve spectrum.  

\begin{figure*}
    \centering
    \plottwo{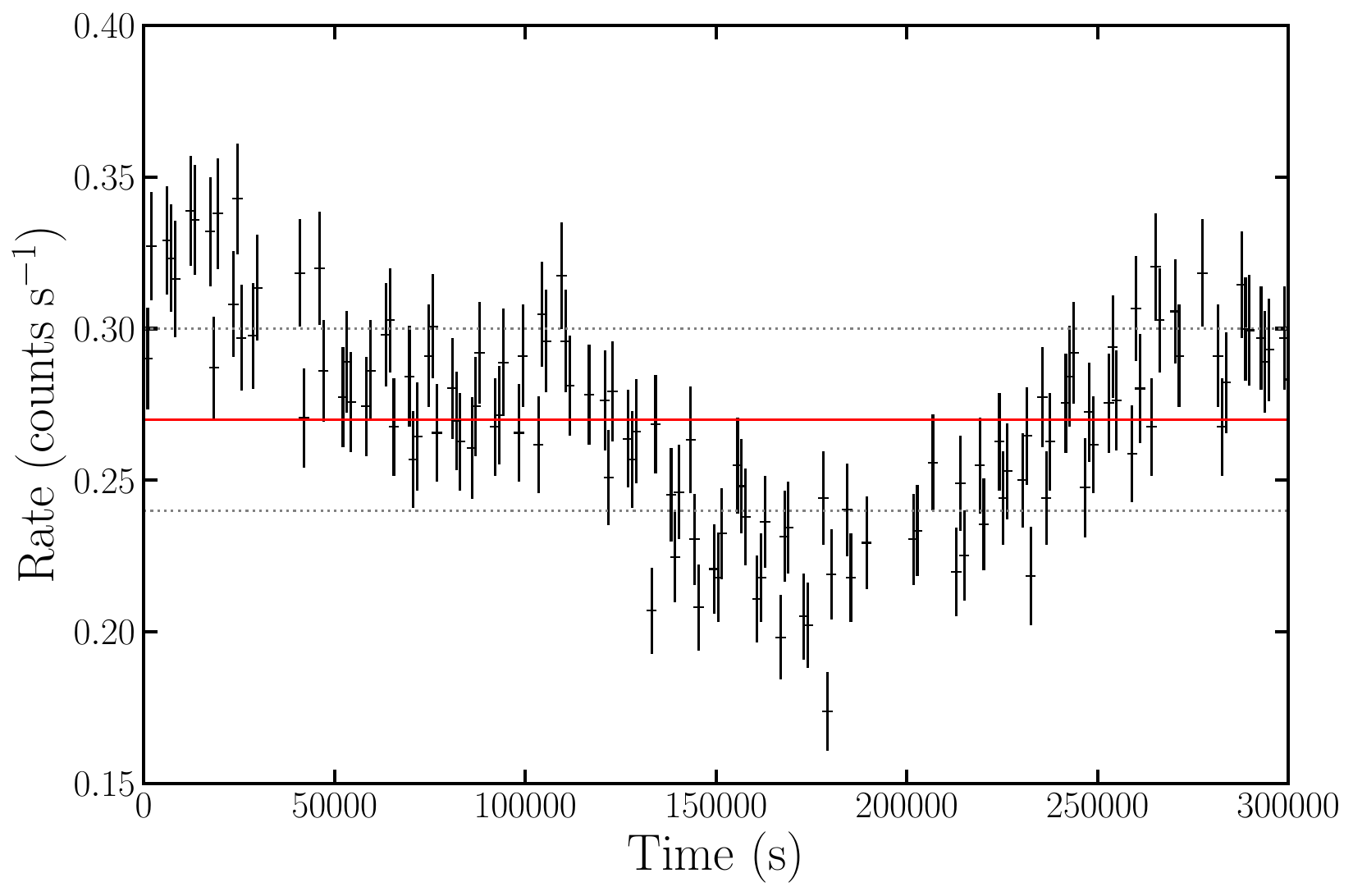}{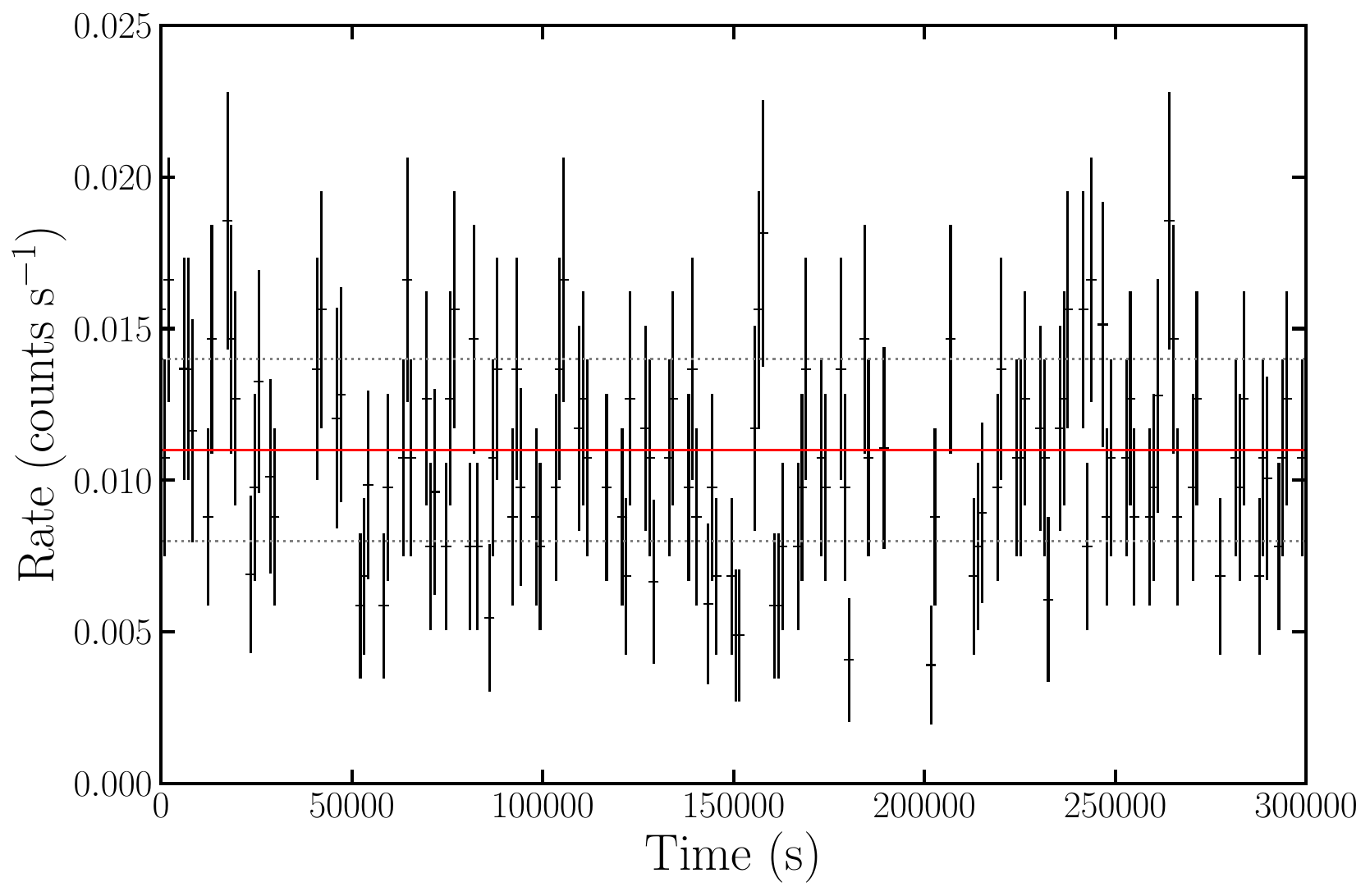}
    \caption{Lightcurves of the XRISM/Resolve observation of Mrk 279. The left panel shows the lightcurve of the full Resolve band, and the right panel shows the lightcurve of the filtered 6.3--6.5 keV band (6.1--6.3~keV in the observed frame). The red horizontal line in each panel indicates the mean rate in each band ($0.27\pm0.03\text{ counts s}^{-1}$ for the full band, $0.011\pm0.003\text{ counts s}^{-1}$ for 6.1--6.3 keV), and the horizontal dotted lines indicate the rms of the variations in each curve.}
    \label{fig:lightcurve_comparison}
\end{figure*}

When XRISM observed Mrk~279, the gate valve was in a closed position,
truncating the spectrum below 1.6~keV, and reducing the total
effective area across the pass band.  The data were reduced following
the XRISM Quick-Start Guide v2.3, via HEASOFT version 6.34 and the
associated calibration files and tools.  Briefly, the data were
filtered to exclude anomalies in low-Earth orbit, and events from the
calibration pixel (36) and pixels 11 and 27 were excluded.  The data
were further screened to include only high-resolution primary (or, Hp)
events, which have a resolution of approximately 4.5~eV
\citep{2024ApJ...973L..25X}.  The light curve of the full Resolve
  band shows only modest variability (see Figure
  \ref{fig:lightcurve_comparison}, left panel), while the light curve
  of the Fe~K band is entirely consistent with a constant count rate
  (see Figure \ref{fig:lightcurve_comparison}, right panel).  The
resulting spectrum is constructed with default bins of 0.5~eV, and the
tools \texttt{rslmkrmf} and \texttt{xaarfgen} are used to make the
redistribution matrix file (rmf) and ancillary response file (arf).
We elected to create a ``Large'' rmf.  After all screening, the net
Resolve exposure time was 168~ks.

NuSTAR and XMM-Newton made contemporaneous observations of Mrk~279 in
support of this XRISM effort.  A joint analysis of the data from all
three observatories will be presented in a forthcoming paper (Byun et
al.\ 2025).

\section{Reduction and Analysis}

All spectral modeling was done using SPEX version 3.08.01
\citep{kaastra1996}, minimizing a Cash statistic
\citep{1979ApJ...228..939C}.  Within SPEX, the data were binned prior
to fitting, using the ``optimal'' binning algorithm of
\cite{kaastra2016}.  To boost the sensitivity of the data, the data
were further binned by a uniform factor of 4.0.  The XRISM/Resolve
spectrum was fit over the 2.0--12.0~keV band to avoid calibration
uncertainties at low energy and to avoid background effects at high
energy.

Following \cite{2024ApJ...973L..25X} we adopted a model with a
blackbody peaking in the UV, and an attenuated power-law continuum in
the X-ray band.  XRISM observations of the Seyfert-1 galaxies NGC~4151
and NGC~3783 both require multiple velocity components to describe the
Fe~K$_{\alpha}$ emission lines observed in these systems
(\citealt{2024ApJ...973L..25X}, \citealt{mehdipour2025}); accordingly,
we allowed for up to three velocity components.  Finally, we searched
for photoionized absorption and emission components, following the
detection of a broad range of outflows in NGC 4151 \citep{xiang2025}
and NGC~3783 \citep{mehdipour2025}, and extremely fast flows in
PDS~456 \citep{pds456}.

\subsection{Basic Emission Model}
\label{subsec:basicmodel}
To provide a relatively model-independent characterization of the
Fe~K$_{\alpha}$ line complex in Mrk~279, we first consider only a
basic model with an attenuated power-law X-ray continuum and emission
line components.  We use the ``pow'' component within SPEX, plus two
``etau'' components to bend the function to zero flux at low and high
energy.  Although many variations on a broken power-law are possible
via ``pow'', only a single power-law index, $\Gamma$, and flux
normalization are allowed to vary.  At calorimeter resolution, the two
components of the Fe~K$_{\alpha}$ line become evident, and a simple
line core is inadequate.  To account for this atomic structure, we use
the ``mytorus'' line model \citep{2009MNRAS.397.1549M}.  This model
has been employed to fit Resolve spectra from NGC 4151, Cen A, and
M81* (\citealt{2024ApJ...973L..25X}, \citealt{bogensberger2025},
\citealt{miller2025}).

The mytorus model parameters include the line flux normalization, the
power-law index of the continuum (coupled to that of the ``pow''
component), the inclination of the emitting gas, a velocity shift
parameter, and the column density of the emitting gas.  Absent an
inclination-sensitive convolution function, the inclination does not
shape the line, and we fix an arbitrary value of 30~degrees.  We found
no evidence of a velocity shift, and fix a value of zero in all
components for convenience.  The data lack the sensitivity to place
strong constraints on the gas column density, so we fix a value of
${\rm N}_{\rm H} = 1.6\times 10^{23}~{\rm cm}^{-2}$, consistent with
values measured in NGC~4151 \citep{2024ApJ...973L..25X}.  Each
``mytorus'' component is broadened with a Gaussian via the ``vgau''
function within SPEX, for which the Gaussian width $\sigma$ is the
only parameter.  Each line component is therefore characterized by a
flux normalization and width.

\begin{figure*}
    \centering
    \includegraphics[width=0.7\textwidth]{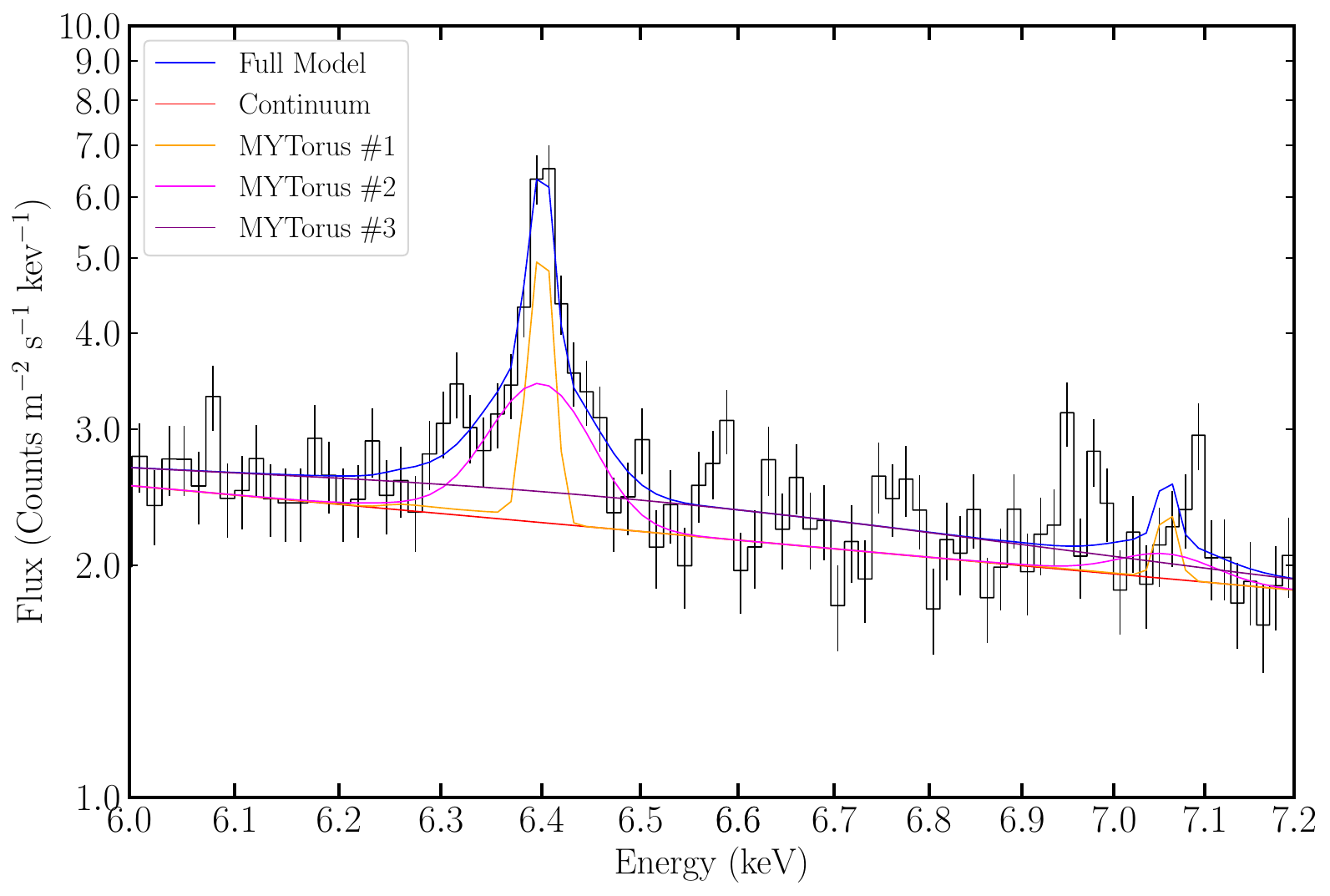}
    \caption{Simple model of Mrk 279 with Gaussian broadened ``mytorus" components. The full model is plotted in blue, while the continuum is shown in red. The individual ``MYTorus" components are shown in orange, magenta, and purple, respectively.  The data are plotted in the rest frame.}
    \label{fig:simplefit}
\end{figure*}

The best fitting basic model parameters are shown in Table
\ref{table:simpleparams}, under the column labeled ``Time-averaged''.
The normalization of the ``mytorus'' components have a consistent
order of magnitude of $\sim10^8$ (in the same units as the power-law),
with a factor of $\sim3$ difference between the smallest and largest
values.  In contrast, the Gaussian widths show larger differences from
one component to the next.  At $\sigma = 370_{-170}^{+150}$, $\sigma =
2200_{-500}^{+600}$, and $\sigma = 20,000_{-4000}^{+6000}\text{ km
  s}^{-1}$, each component is separated by an order of magnitude
difference. The individual components as modeled are shown in Figure
\ref{fig:simplefit}.

\subsection{A Detailed Physical Model}

To better understand the geometry of the accretion flow, we replace
simple Gaussian broadening of the Fe~K$_{\alpha}$ emission line
components with the SPEX function ``spei.''  This function includes
relativistic effects and is sensitive to the inclination of the
emission region(s).  Owing to the limited sensitivity of the data, we
hold most parameters fixed at their default values, and elect to fit
for the inner emission radius and inclination of each component.  Of
the fixed values, the most important is the emissivity index ($J
\propto r^{-q})$.  We adopt a value of $q=3$ as per a flat disk; in
this case, the measured inclinations may better reflect the difference
between the systemic inclination and the opening angle of the emission
geometry relative to a planar disk \citep{2024ApJ...973L..25X}.

We search for photoionized emission and absorption using the ``pion''
model.  This model calculates the emission and absorption from a slab
of gas, using the best-fit continuum at each step of the minimization
process.  The spectral energy distribution of AGN is dominated by UV
or EUV emission; obtaining robust results from pion requires an
accurate characterization of the SED over the 0.0136--13.6~keV band.
We examined archival Hubble UV spectra of Mrk~279, but we were not
able to derive a reliable continuum at the shortest, most critical
wavelengths.  Broadly following the procedure adopted in NGC 4151
\citep{xiang2025}, we include a blackbody with a temperature of
$k\,T=0.017$~keV with a 10:1 flux ratio relative to the weaker
power-law.  Both of these continuum components are then fed into
``pion.''

The pion absorption components are characterized by fitting for the
gas column density (${\rm N}_{\rm H}$), ionization parameter ($\xi =
L/nr^{2}$, where $n$ is the gas number density and $L$ is the ionizing
luminosity) velocity broadening ($\sigma_v$), and bulk velocity shift
($v_z$).  For simplicity, we assume a covering factor of 0.5 for all
absorption components.  Rather than random searches for winds, we
conduct limited, directed searches for absorption from ``warm
absorbers'' at very low shifts, and flows with velocities close to $v
\simeq 0.3c$ as per PDS~456 \citep{pds456}.  In practice, column
density and the geometric covering factor ($f_{\rm cov} =
\Omega/4\pi$) of emission components are partly degenerate for
photoionized emission.  In searching for photoionized emission at low
velocity shifts, then, we fix an optically thin column density of
$N_{\rm H} = 1\times 10^{23}~{\rm cm}^{-2}$ and fit for the covering
factor in a range of 0 to 1.  The total best-fit physical model can be
summarized as follows:
\begin{multline}
    reds*hot*[(bb + pow * etau_{low} * etau_{hi} + MYT_1*spei\\ + MYT_2 *spei + MYT_3*spei)*pion_1*pion_2*pion_{3}] \\
\end{multline}

Here, pion$_1$ and pion$_2$ model absorption, and pion$_3$ models
emission.  The Cash statistic of the best-fit physical model is
$C=693$ for $\nu=664$ degrees of freedom, well within the expected
Cash statistic of $690 \pm 37$, indicating a good fit
\citep{Kaastra_2017}. Table \ref{table:physical_parameters} presents
the best-fit parameters for the physical model, as well as some
derived properties for the absorption components, including launching
radii, mass outflow rates, and kinetic powers. The fit in a broad band
view is shown in Figure \ref{fig:physical_brd}. Figure
\ref{fig:physical_comp} depicts the breakdown of each emission and
absorption component of the best-fit model.  The best-fit model gives
X-ray flux of $F_{2-10} = 2.1
\times10^{-11}~\mathrm{erg~cm^{-2}~s^{-1}}$ in the 2-10 keV band and a
total ionizing luminosity of $L_{\rm ion}^{\rm all} = 1.3 \times
10^{45}~\mathrm{erg~s^{-1}}$. Assuming $L_{\rm bol} = 2L_{\rm
  ion}^{\rm all}$, we estimate the bolometric luminosity of Mrk 279 to
be approximately $2.6 \times 10^{45}~\mathrm{erg~s^{-1}}$, yielding an
Eddington ratio of $\lambda_{\it Edd} = L_{\rm bol}/L_{\rm Edd} = 0.5$
with $L_\mathrm{\rm Edd} \equiv 4\pi GM_{\rm BH}m_{\rm p}c/\sigma_{\rm
  T} \simeq 4.8\times 10^{45}~\mathrm{erg~s^{-1}}$.

The significance of each Fe~K$_{\alpha}$ emission line component and
each pion component is evaluated by removing it from the full model,
refitting, and examining the difference in the fit statistic.  To
assess the significance of each component, we compute the Small Sample
Akaike Information Criterion (AIC) \citep{Akaike_1974,
  Emmanoulopoulos_2016}, defined as $\rm AIC = 2\,p - 2\,C_L + C +
2\,p(p+1)/(n-p-1)$, where $C$ is the C-stat of the fit, $C_L$ is the
C-stat of the best fit model, $n = 687$ is the number of data bins,
and $p = 23$ is the number of free parameters in the model. In
Table~\ref{table:physical_parameters}, we report the change in AIC
($\Delta \rm AIC$).  A value of $\Delta \rm AIC < -10$ indicates that
the component is strongly required to account for the spectral
features, while $\Delta \rm AIC < -2$ provides substantial evidence
\citep{Burnham_2002}.  We also report a complementary detection
significance (D.S.) in units of Gaussian-equivalent sigma, derived
from the p-value associated with the changes in C-stat and the
difference in degrees of freedom.

\subsubsection{Neutral Fe K$\alpha$ emission components}
The significance of each line component within the total model is
listed in Table 2.  The narrowest component is the most significant,
followed by the intermediate-width component, and the broadest
component.  The broadest component is clearly preferred by the model
as judged by the AIC, but is only significant at the $2.4\sigma$ level
of confidence when only considering the p-value associated with
changes in the Cash statistic.  These tests measure the significance
of adding a third component to a complex line profile.  A full measure
of the significance of each component might be better indicated by
dividing the flux of each component by its $1\sigma$ minus-side error.
This metric indicates that the components are significant at the
$5\sigma$, $4.3\sigma$, and $3\sigma$ levels of confidence.

The narrow core of the Fe~K$_{\alpha}$ line is found to originate at a
radius of $r = 22000^{+231000}_{-11000}~GM/c^{2}$.  This is likely
better viewed as a lower limit of $r \geq 10^{4}~GM/c^{2}$.  These
values are compatible with the inner wall of the molecular torus.  The
intermediate-width component is measured to originate at $r =
1560^{+5750}_{-1480}~GM/c^{2}$.  This range of radii is compatible
with an inner X-ray extension to the optical BLR.  The difference in
implied radius between this component and the narrow core is
consistent with dust reverberation studies that establish the inner
wall of the torus is only a few times larger than the optical BLR
\citep{minezaki2019}.  The broadest component of the emission line is
found to have an inner radius of $r = 81^{+280}_{-75}~GM/c^{2}$.  It
is important to note that this value includes the innermost stable
circular orbit for a black hole with low spin, but would also allow
for a radially truncated disk. 

The inclinations of all three components are statistically consistent,
but poorly constrained.  The upper limits allow for inclinations
exceeding 80~degrees, which is implausible in an unobscured Seyfert-1
AGN.  The large uncertainties on the inclination are a reflection of
the lower flux and modest Fe~K$_{\alpha}$ emission line observed from
Mrk~279, and they are partly responsible for large uncertainties on
the inner radius of each component.

\subsubsection{Photoionized emission and absorption components}
\begin{figure*}
    \centering
     \includegraphics[width=0.9\textwidth]{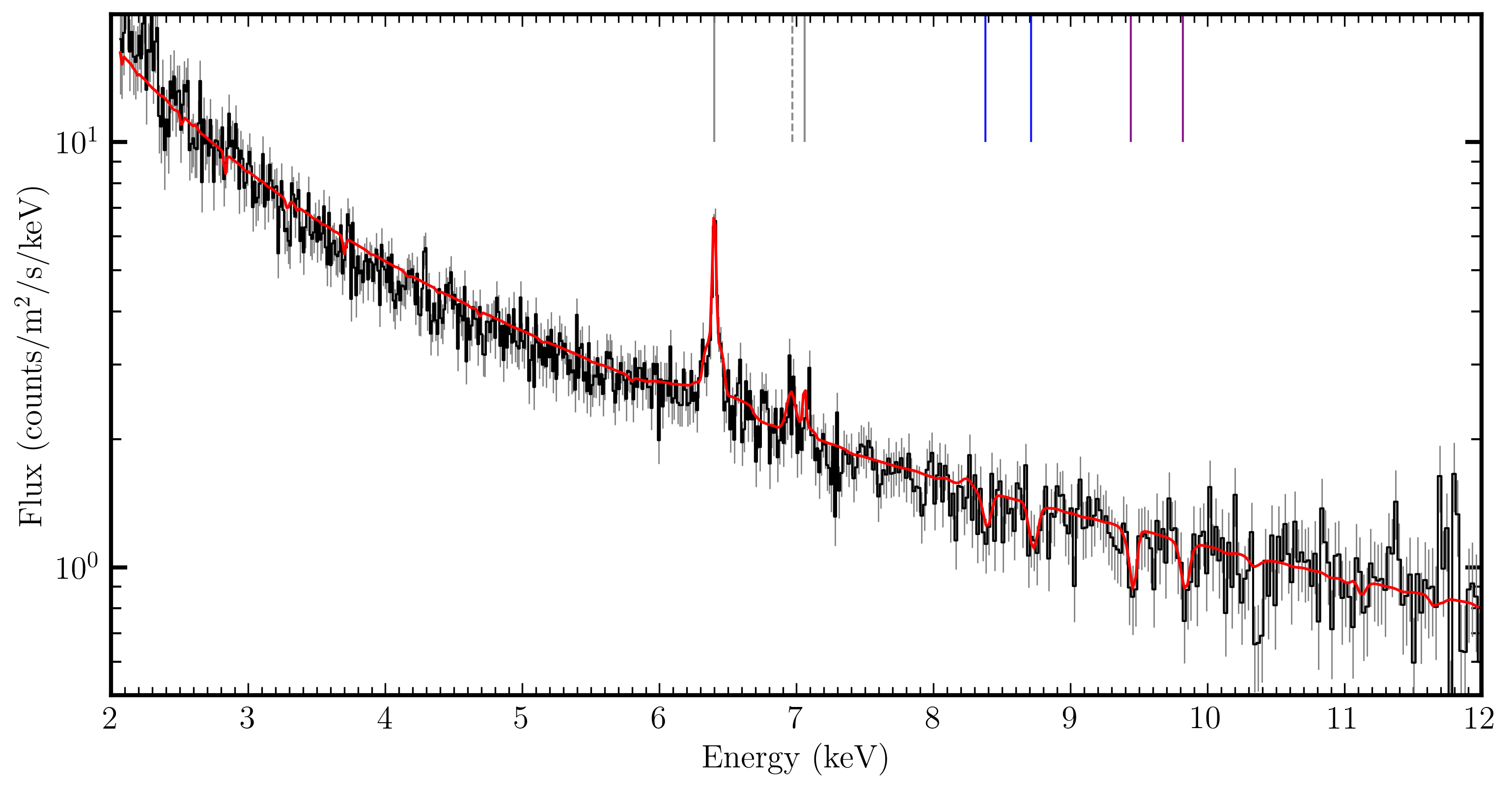}
   \caption{The time-averaged XRISM/Resolve spectrum of Mrk 279.  The data are plotted in the rest frame, and binned using the ``optimal'' algorithm for fitting, and by a further factor of 4.0.  The model shown in red is the best-fit extended physical model, including three components for the neutral Fe~K emission lines, photoionized emission contributing an H-like Fe~XXVI line at 6.97~keV, and photoionized absorption from tentative fast wind components.  The solid gray lines indicate the lab energies of the neutral Fe~K$_{\alpha}$ and Fe~K$_{\beta}$ lines.  The dashed gray line indicates the lab energy for the Fe~XXVI emission line.  The blue and purple lines indicate He-like Fe XXV and H-like Fe XXVI line pairs from potential ultra-fast outflows at $v\simeq 0.22c$ and $v\simeq 0.33c$.  Please see the text and Table 2.}
   \label{fig:physical_brd}
\end{figure*}

\begin{figure*}
    \centering
     \includegraphics[width=0.9\textwidth]{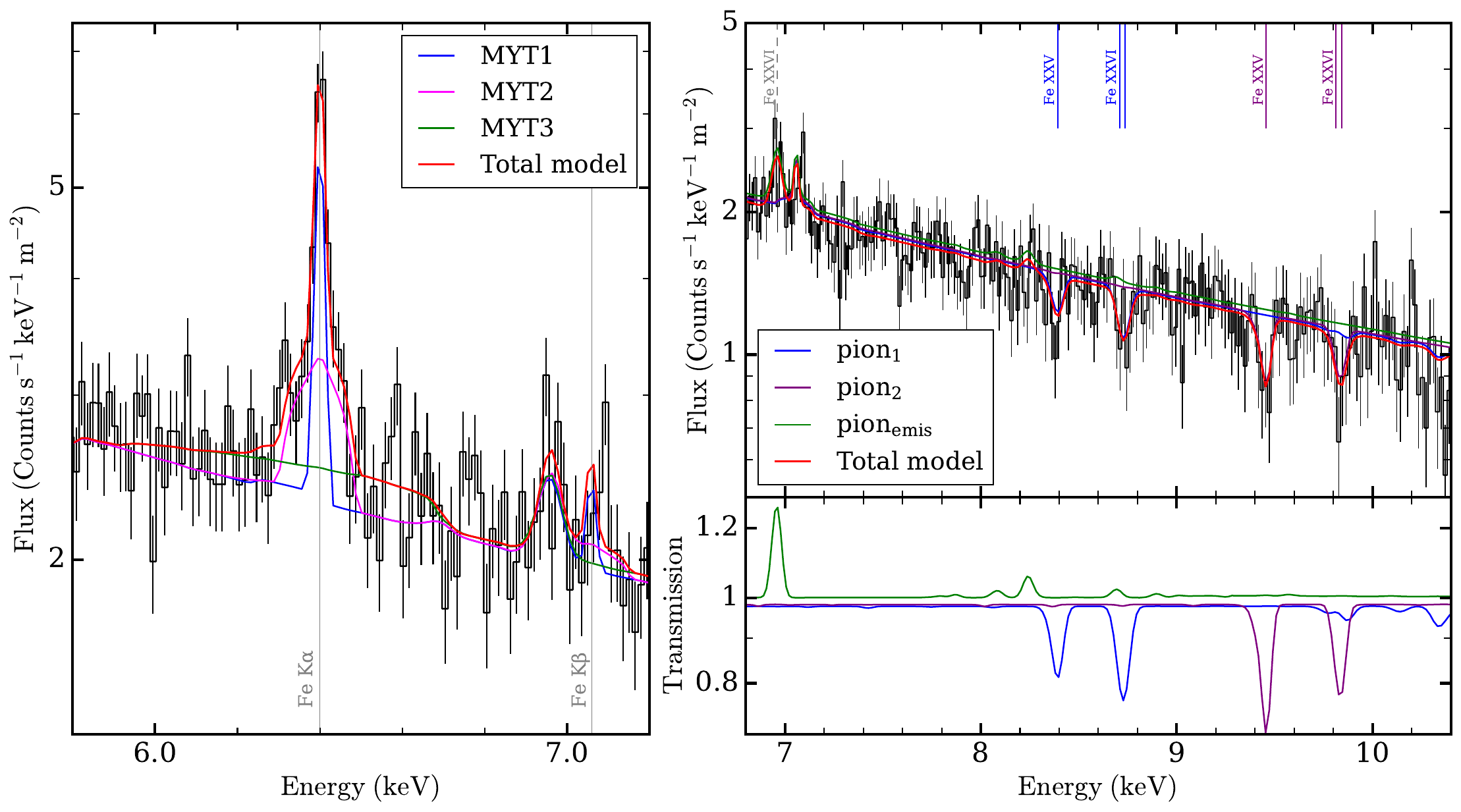}
   \caption{The zoomed-in spectrum and the best-fit extended physical model (red), including the breakdown of the components for Fe~K emission lines and photionized emission/absorption lines.  The data are plotted in the rest frame.}
   \label{fig:physical_comp}
\end{figure*}

The fits revealed one photoionized emission component (${\rm
  pion}_{\rm 3}$).  The appearance of Fe~XXVI emission
(Ly$\alpha_1$ and $\alpha_2$ at 6.95 and 6.97~keV) and the absence of
Fe~XXV emission (close to 6.7~keV) leads to a very high ionization
parameter, ${\rm log}~{\xi} = 4.8 \pm 0.1$ (the ionization parameter
$\xi$ has units of ${\rm erg}~{\rm cm}~{\rm s}^{-1}$).  The covering
factor hit the upper bound of 1.0, with a $1\sigma$ lower limit of
$0.58$; this signifies that the emitting gas may have a nearly
spherical distribution.  This component is nominally but not
significantly red-shifted, with a velocity of $v_z =
360^{+390}_{-370}~{\rm km}~{\rm s}^{-1}$.  Factoring in all of these
properties, the emission may be a slow, wide-angle wind or failed
wind.  For this gas not to appear as a ``warm absorber'', its column
density must be particularly low, or it must lie above the direct line
of sight to the central engine.  Its ionization is much higher than
X-ray components that have been observed in the optical ``narrow-line
region'' (or, NLR; see \citealt{wang2011}).  The line position is
marked as grey dashed vertical lines in both Figure
\ref{fig:physical_brd} and Figure \ref{fig:physical_comp}.  The
emission component is statistically required with $\Delta {\rm AIC} =
-8.3$ (DS = $3.1 \sigma$), and appears to confirm prior reports of
Fe~XXVI emission in Mrk~279 (e.g., \citealt{costantini2007},
\citealt{costantini2010}).

Two absorption components (${\rm pion}_1$ and ${\rm pion}_2$) are
tentatively detected, with blue-shifted velocities of approximately
$v_z \sim 6.6\times10^{4}~\mathrm{km~s^{-1}}$ and $v_z \sim
9.9\times10^{4}~\mathrm{km~s^{-1}}$ or $0.22c$ and $0.33c$ (see Figure
\ref{fig:physical_brd} and Figure \ref{fig:physical_comp}).  These
components represent only modest improvements to the fit, with $\Delta
{\rm AIC} = -3.3$ and $-3.5$, respectively (roughly $2.4\sigma$
individually), and we regard them with some skepticism.  The
  narrow widths of the lines in these zones ($\sigma =
  920^{+260}_{-590}~\mathrm{km~s^{-1}}$ and $\sigma =
  670^{+200}_{-240}~\mathrm{km~s^{-1}}$, respectively) would require
  clumping in order for the velocities to remain distinct through an
  acceleration phase.  A brief exploration of the feedback properties
  of these tentative outflows is merited.  In this examination, we
  broadly follow the procedure in \cite{xiang2025}.

The mass outflow rate can be written as:
\begin{equation} \label{eq:M_dot}
    \dot M_{\mathrm{out}} = 4 \pi f_{\mathrm{cov}}\mu m_p \frac{L_{\rm ion}^{\rm in}}{\xi}v_{\mathrm{\rm out}}f_\mathrm{v}~,
\end{equation}
where $\mu = 1.23$ is the mean atomic weight, $m_p$ is the proton
mass, $f_{\rm cov}$ and $f_\mathrm{v}$ are angular and volume filling
factors, and we have used the ionization parameter formalism to
replace $r^{2}$ with $L_{\rm ion}/\xi$.  Within SPEX, the ionizing
luminosity incident on each absorber $L_{\rm ion}^{\rm in}$ is
determined directly from the outputs parameter ``lixi'' of ``pion''
component, defined as $L_{\rm ion}^{\rm in}/{\xi}$.

Assuming a unity filling factor $f_\mathrm{v}=1$ (the upper limit),
the mass outflow rates for the two absorbers are $\dot{M} =
138^{+81}_{-72}~{\rm M}_{\odot}~{\rm yr}^{-1}$ and $\dot{M} =
300^{+175}_{-101}~{\rm M}_{\odot}~{\rm yr}^{-1}$.  These rates far
exceed the Eddington mass accretion rate $\dot M_\mathrm{Edd} \equiv
L_\mathrm{Edd}/ \eta c^2 = 0.8~(M_\odot / \rm yr)$, for radiative
accretion efficiency of $\eta = 0.1$.

The Kinetic power of a wind, assuming it reaches a terminal velocity without significant acceleration, is given by:
\begin{equation} \label{eq:E_k}
    \dot E_k = \frac{1}{2} \dot M_{\mathrm{out}} v_{\mathrm{out}}^2
\end{equation}
For a unity volume filling factor, the two UFOs have kinetic power of
$L_{\rm kin} = 19^{+11}_{-10} \times 10^{46}$ and $L_{\rm kin} =
93^{+55}_{-32} \times 10^{46}~\mathrm{erg~s^{-1}}$.
These values are approximately $40 \times$ and $200 \times$ more than the
radiative Eddington luminosity of Mrk~279.

The launching radius of the wind can be constrained using two
complementary estimates that provide upper and lower limits, assuming
that the gas originates in a thin shell ($\Delta r/r << 1$).  Assuming
the radial thickness of the absorber ($\Delta r$) does not exceed its
distance from the source, and using the relation
$N_H = n\Delta r \leq f_\mathrm{v}nr$, the upper limit becomes:
\begin{equation} \label{eq: r_max}
    r_{\mathrm{max}} =  \frac{L_\mathrm{ion} f_\mathrm{v}}{\xi N_\mathrm{H}} = r_1 f_\mathrm{v},
\end{equation}
where $f_\mathrm{v}$ is the volume filling factor, and $r_1 =
L_{\mathrm{ion}} / \xi N_{\rm H}$ is the so-called ``absorption
radius'', which can be computed directly from the measured values of
$N_{\rm H}$ and $\xi$. This absorption radius provides a useful upper
bound on the launching radius.

On the other hand, assuming that the observed outflow velocity is
comparable to the local escape velocity, we can estimate a lower limit
for the launching radius, as follows:
\begin{equation} \label{eq: r_min}
    r_{\mathrm{min}} = \frac{2GM}{v^2_{\mathrm{out}}} = \frac{2GM}{v^2_{\mathrm{z}}} \cos^2{\theta} = r_2 \cos^2{\theta},
\end{equation}
where $r_2 = 2GM / v_{\mathrm{z}}^2$ is referred to as the ``wind
radius'', $v_{\mathrm{z}}$ is the line-of-sight velocity, and $\theta$
is the inclination of the flow relative to the line of sight. For
small inclination angles ($\cos \theta \approx 1$), the ratio
$r_2/r_1$ provides an estimate of the minimum volume filling factor,
with the constraint that $r_{\mathrm{max}} \geq r_{\mathrm{min}}$.

The inferred ``absorber radius'' for the two absorption components are
$1.6^{+0.5}_{-0.6} \times 10^5$ and $3.1^{+1.1}_{-0.8} \times 10^5$,
and the ``wind radius'' values are $41^{+0.7}_{-0.3}$ and
$18^{+0.2}_{-0.1}$, in units of $GM/c^2$. The overlapping ranges of
their inferred radii suggest that the two components may have a common
origin. The lower limits for the volume filling factors are
$2.5^{+1.5}_{-0.6} \times 10^{-4}$ and $0.6^{+0.2}_{-0.1} \times
10^{-4}$.  If these values are correct, the mass outflow rates would
be comparable to the inferred accretion rates, and the putative UFOs
in Mrk~279 would plausibly fall below the $L_{\rm kin} \geq 0.5-5\%
L_{\rm Edd}$ threshold for galaxy altering feedback
(\citealt{dimatteo2005}, \citealt{hopkins2010}).

To further investigate the energetics and possible driving mechanisms, we calculate the outflow momentum rate,
\begin{equation} \label{eq:P_out}
    \dot p_{\mathrm{out}} = \dot M_{\mathrm{out}} v_{\mathrm{out}},
\end{equation}
and the radiation momentum flux,
\begin{equation} \label{eq:P_rad}
    \dot p_{\mathrm{rad}} = \frac{L_{\rm bol}^{\rm in}}{c},
\end{equation}
where $L_{bol}^{in}$ is the corrected bolometric luminosity incident on each absorber assuming $L_{\rm bol}^{\rm in} = 2L_{\rm ion}^{\rm in}$. 

Assuming a unity volume filling factor, the two outflow components
carry momentum fluxes of $\dot p_{\mathrm{out}} \sim 10^{37-38}$ dyn,
exceeding by 3--4 orders of magnitude their corresponding radiation
momentum fluxes ($\dot p_{\mathrm{rad}} \sim 10^{34-35}$ dyn). Under
these conditions, radiation pressure alone is insufficient to
accelerate these outflows.  Radiation pressure on lines alone is
likely ruled out by the high ionization of the putative UFO
components, which leaves no UV transitions and provides no force
multiplier.  In this scenario, it is not clear that even magnetic
pressure could drive the flows.

If, however, the absorbing gas is highly clumpy, where the effective
filling factor could be as low as $\sim 10^{-4}$, the outflow momentum
rate drops to only $30\%$ of the radiation momentum flux. In this
case, radiation pressure in the form of electron scattering pressure
could dominate.  In scenarios where electron scattering pressure plays
an important role, a direct proportionality between $\dot
p_{\mathrm{out}}$ and $\dot p_{\mathrm{rad}}$ is expected
\citep{Gofford_2015}. The fact that $pion_2$, which has the larger
outflow momentum, also exhibits the larger radiation momentum flux is
consistent with this picture.

\section{Discussion}
\label{sec:discussion}
We analyzed the first XRISM/Resolve spectrum of Mrk~279, a nearby
Seyfert-1 AGN with a history of compelling properties, including a
complex and variable Fe~K$_{\alpha}$ emission line profile.  Adopting
a set of models similar to those applied to the XRISM spectra of
NGC~4151, we find that the line complex contains components that can
be linked to the inner face of the torus, an inner extension of the
optical BLR, and the inner accretion disk.  In this sense, our
observation confirms that even narrow Fe~K$_{\alpha}$ lines represent
incisive probes of the central engine in Seyfert galaxies.  Similar to
prior studies, the Resolve spectrum does not require slow, ionized,
``warm absorber'' winds in the Fe~K band.  The data reveal tentative
indications of a very fast wind with at least superficial similarities
to the UFOs that XRISM has revealed in NGC~4151 and especially PDS~456
(\citealt{xiang2025}, \citealt{pds456}).  In this section, we place
the geometric constraints obtained from Mrk~279 into the broader
context of recent XRISM results, critically examine the putative fast
outflows and how they might be drive, and suggest directions for
future studies of Mrk~279.

\subsection{Fe~K lines and accretion flow geometries}
XRISM observations of NGC~4151 established that the components of the
Fe~K$_{\alpha}$ emission complex may trace all of the key geometries
within the central engine \citep{2024ApJ...973L..25X}.  Our
observation of Mrk~279 appears to confirm this, and signals that
similar insights can be gained even from AGN with lower fluxes.  The
net exposure of 168~ks required 300~ks or 3.5~days to accumulate,
owing to the modest efficiency of low-Earth orbit.  This is much
shorter than the H$\beta$ reverberation lag timescale of $\tau = 18\pm
1$~days \citep{Zu_2011}.  Our single XRISM observation does not permit
the detection of a clear line response to time variations in the
central engine.

The common features of the spectra in Mrk~279 and NGC~4151 are at
least qualitatively consistent with a picture wherein the intermediate
component the narrow Fe~K line traces an inner, low-velocity region of
the BLR close to the disk.  If the BLR represents an outflow, the
  gas must be detected before acceleration occurs downstream.  This
may offer qualitative support for models that rely on dust -- onto
which iron may accumulate -- to launch the winds that we identify as
the BLR \citep{czerny2015}.  If the putative UFOs are confirmed in
later work (see below), the broadest emission component could
plausibly be associated with emission from the wind, similar to the
case of PDS~456 \citep{pds456}.

\begin{figure*}
    \centering
    \includegraphics[width=0.9\textwidth]{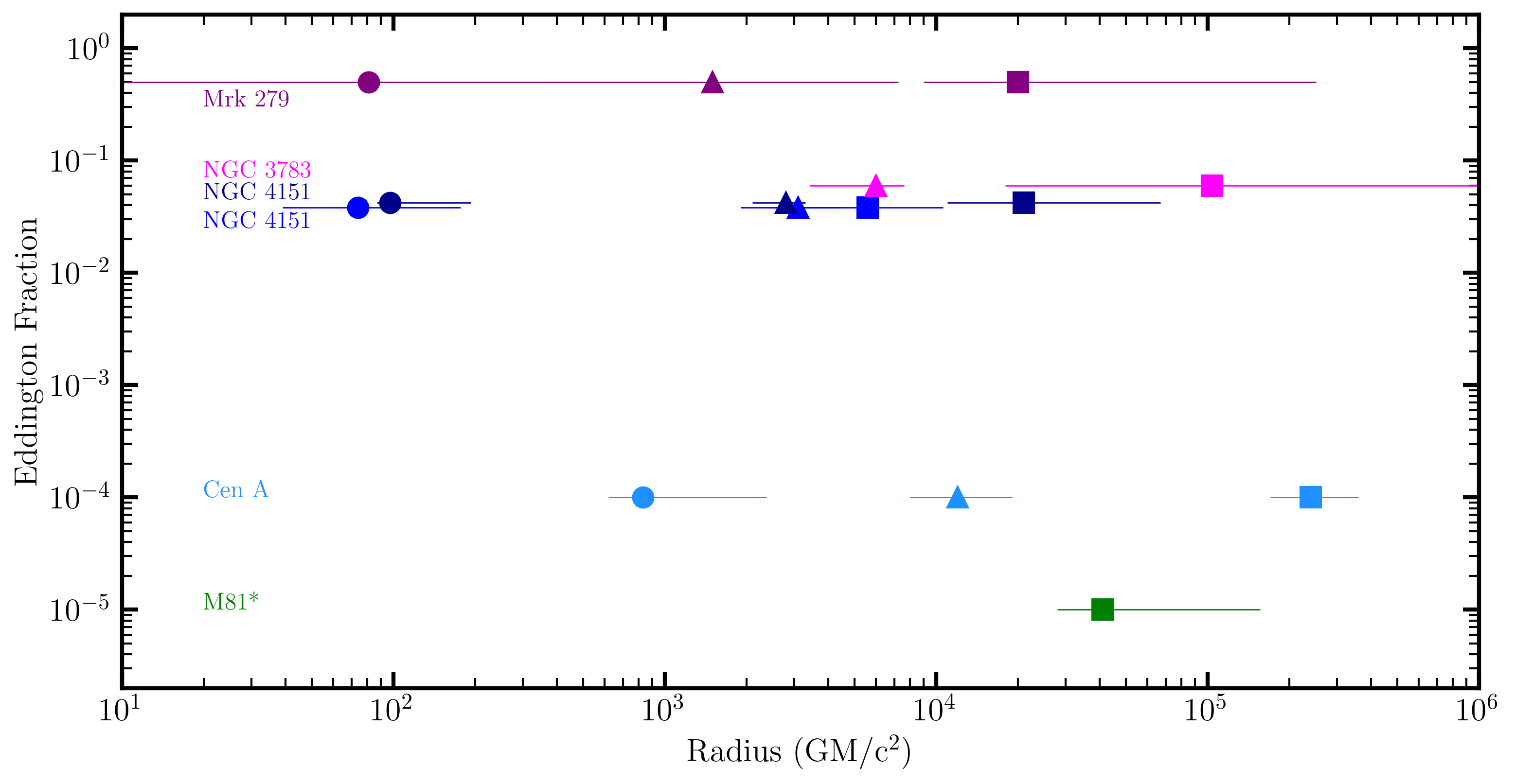}
    \caption{A comparison of the inner radii measured for Fe~K$_{\alpha}$ line components in a subset of XRISM spectra of local AGN.  All of the radii are based on blurred ``mytorus'' modeling of Resolve spectra; please see the text for details.  With several important caveats, this plot suggests that Resolve data are able to separate line emission from the inner disk, BLR, and torus, and capable of tracing trends with Eddington fraction.}
    \label{fig:eddrad}
\end{figure*}

The efficacy with which X-rays reveal the geometry of the central
engine should also be assessed by looking at a broad range of
sources, and examining potential trends with parameters like the
Eddington fraction.  We therefore collected radius constraints from
the recent literature on XRISM observations of AGN.  Similar to our
analysis, fits with Speith-broadened ``mytorus'' components have been
reported from NGC~4151, Cen~A, and M81*
(\citealt{2024ApJ...973L..25X}, \citealt{bogensberger2025},
\citealt{miller2025}).  To these, we added the corresponding results
from a forthcoming work on the first XRISM/Resolve spectrum of the
Seyfert-1 galaxy NGC~3783 (Li et al.\ 2025, in prep.),

Figure \ref{fig:eddrad} plots the approximate Eddington fraction of
each source versus the inner radius of the components measured in
Resolve spectra.  (If the Eddington fraction drives the nature of the
accretion flow, it is really the independent variable, but the plot
depicts radius on the horizontal axis because this preserves notional
contact with cartoons for how different geometries are laid-out with
radius.)  Although the uncertainties on the components in Mrk~279 are
larger than for NGC~4151, it is clear that the inferred radii are in
formal agreement.  Here, it is again worth noting that some of the
inclinations are poorly constrained, and that the results depend on
the assumed emissivity profile, but in this comparison the same
emissivity is assumed for all sources.  The line complex in NGC 3783
includes a very broad line component from the innermost disk,
potentially as close as $2~GM/c^{2}$ (Li et al.\ 2025).  The line in
M81* does not statistically require multiple components; its single
component width is consistent with the torus values inferred in NGC
4151 and Mrk 279.

The inclination of the optical BLR in Mrk~279 is constrained to be
$\theta = 29.1\pm 3.4$~degrees with respect to our line of sight
\citep{williams2018}.  This is consistent with the value that we have
measured from the intermediate-width component of the Fe~K$_{\alpha}$
line, $\theta = 34^{+53}_{-24}$~degrees, although our uncertainties do
not provide a strong constraint.  In some AGN, the broad accretion
flow geometry may be consistent with a ``bowl'' geometry (e.g.,
\citealt{gravity2024}); in NGC~4151, the observation of progressively
lower effective inclinations with radius may indicate such a geometry
\citep{2024ApJ...973L..25X}.  Deeper XRISM observations of Mrk~279
that achieve tighter inclination constraints will be able to test
this.

Our three-component model for the Fe~K emission lines may not capture
all of the information in the Mrk~279 spectrum, or it may fail to
interpret all of the information correctly:

First, the Fe~K$_{\alpha}$ line complex may have an
intriguing degree of complexity on its red wing (see Figure 2).  In
the time-averaged and flux-selected spectra, there is evidence of
narrow emission at 6.32~keV.  This putative feature is inconsistent
with the edge expected at 6.25~keV from 150~eV losses in 180-degree
Compton scattering.  We find that it can be modeled in terms of a very
narrow ring of emission, only tens of radii from the black hole.
Qualitatively, this may be consistent with numerical simulations of
warped and torn disks in accretion disks that are misaligned with the
black hole spin vector (e.g., \citealt{nixon2012}).  However, this is
an extreme interpretation, and deeper observations are required to
test it.

Second, our model may be deficient in assuming that
all of the K$_{\alpha}$ and K$_{\beta}$ flux is due to neutral Fe.
The predicted K$_{\beta}$ line is only just consistent with the
$1\sigma$ errors on the data (see Figures 1 and 2).  Moreover, at only
a slightly higher energy, just shy of 7.1~keV, a bin stands above the
continuum.  This may signal a contribution from slightly ionized iron.
In this case, the blue wing of the K$_{\alpha}$ complex would also
have to correspond to ionized iron, rather than just velocity
broadening of a neutral line.  The data disfavor this interpretation
because a good fit to the the K$_{\beta}$ line would require a
K$_{\alpha}$ line that is stronger than the data.  It is possible that
the appearance of the Fe~K$_{\beta}$ line is impacted by a $v\simeq
4000~{\rm km}~{\rm s}^{-1}$ wind that covers or distorts this emission
with Fe~XXVI absorption.

We also note that a small upturn in the data below 3~keV causes our
model to under-estimate the data there, and to slightly over-estimate
the data in the 3--4~keV band.  This disparity is likely the result of
our use of the large or ``L'' format response matrix, rather than the
extra-large of an ``XL'' matrix.  This was merely an expedient choice,
because fits and error scans with complex convolution functions {\it
  and} layered photoionization models are extremely difficult with
``XL'' matrices.  We conducted additional fits with a broken power-law
to understand the influence of the artificial upturn at low energy;
the fit through the Fe~K band and above is unaffected.

\subsection{The Search for UFOs}
Our targeted search for UFOs in Mrk~279 was motivated by its
relatively high Eddington fraction, and the dramatic XRISM spectra
obtained from PDS~456 \citep{pds456}.  In that source, the narrow
nature of the observed lines nominally points to confined clumps of
gas.  If the same arguments regarding volume filling factors and clump
sizes apply in Mrk~279, the fast winds would transfer hundreds of
solar masses of material per year and enormous kinetic energy, easily
meeting the theoretical threshold for galaxy shaping feedback
(\citealt{dimatteo2005}, \citealt{Hopkins_2010}).  However, unlike
PDS~456, there is no obvious flaring in the light curve of Mrk~279 to
enable constraints on distances, size scales, and filling factors.  It
is possible that the putative UFOs in Mrk~279 -- if real -- have a
much lower volume filling factor, and therefore convey orders of
magnitude lower mass flux and kinetic power.  This may be more
plausible given that the kinetic power of a true PDS~456-like wind
would exceed the radiative Eddington limit in Mrk~279 by two orders of
magnitude.

Deeper XRISM observations of Mrk~279 -- for instance, with the gate
valve in the ``open position'' -- will be able to test these weak
indications for UFOs.  In this configuration, H-like lines from low-Z
elements that are predicted by the ``pion'' models in Table 2 will be
within the pass band, and should be resolved and detected.  It may
also be possible to target future observations of Mrk~279 to observe a
particularly bright X-ray phase, an interval wherein the emission is
soft and less likely to fully ionize the same fraction of the gas in
our line of sight, or other favorable circumstances.  A promising
approach might be to use variability in the absorption features to
infer the gas density; this would demand a series of coordinated,
timed observations.

The Resolve spectrum of Mrk~279 does not require a ``soft excess''
component owing to its limited pass band.  However, prior XMM-Newton
observations require such a component at low energy
\citep{costantini2010}.  For plausible characterizations of the soft
excess in terms of a steeper power-law or cool thermal component and
plausible flux levels, we find that including this component in the
SED only affects column densities and ionization parameters at the few
percent level.  Velocity widths and shifts are unaffected.  The nature
of the soft excess could also be investigated in an observation with
the gate valve in an open position.

It is worth remarking that both Mrk~279 and PDS~456 lack evidence of
slow-moving, highly ionized ``warm absorber'' winds in the Fe~K band.
In contrast, spectra of NGC~4151 reveal a rich diversity of wind
speeds, including these extremes.  It is tempting to speculate, then,
that the most ionized warm absorbers may be suppressed at high
Eddington fractions.  Narrow-line Seyfert-1 AGN are thought to harbor
relatively low-mass black holes accreting at high Eddington fractions;
of these, NGC~4051 is the most proximal and brightest case. Even in
the deepest Chandra gratings spectra of this source, there is little
evidence for slow winds evident in the Fe~K band
\citep{ogorzalek2022}.  If future XRISM observations of a broader
array of AGN confirm this, it may indicate that faster, wider-angle
UFOs launched at higher Eddington fractions act to sweep away gas that
was previously seen as highly ionized warm absorbers.

We thank the anonymous referee for insights that improved this
manuscritpt.  JMM acknowledges helpful conversations with Liyi Gu and
Jelle Kaastra.

\software{astropy \citep{astropy:2013,astropy:2018,astropy:2022}, SPEX \citep[v3.08.01,][]{kaastra_2024_12771915}, Matplotlib \citep{Hunter:2007}, NumPy \citep{harris2020array}, pandas \citep{mckinney-proc-scipy-2010,the_pandas_development_team_2024_13819579}}

\bibliography{main}{}
\bibliographystyle{aasjournalv7}

\begin{deluxetable}{lcccc}
    \tablenum{1}
    \tablecaption{Basic Model Parameters\label{table:simpleparams}}
    \tablehead{\colhead{Component}&\colhead{Parameter}&\colhead{Value}}
    \startdata
    \text{Power Law} & $\Gamma$& $1.68_{-0.01}^{+0.01}$ \\
    &$A (10^{44}\text{ ph s}^{-1}\text{ keV}^{-1})$ & $1.06_{-0.02}^{+0.02}\times10^8$ \\
    \hline
    \text{MYTorus \#1} &Normalization&$9.2_{-2.5}^{+2.5}\times10^7$\\
    &$\sigma_v$ (km s$^{-1}$)&$370_{-170}^{+150}$\\
    \hline
    \text{MYTorus \#2} &Normalization&$1.5_{-0.3}^{+0.3}\times10^8$\\
    &$\sigma_v$ (km s$^{-1}$)&$2200_{-500}^{+600}$\\
    \hline
    \text{MYTorus \#3} &Normalization&$2.2_{-0.6}^{+0.6}\times10^8$\\
    &$\sigma_v$ (km s$^{-1}$)&$20^{+6}_{-4}\times10^{3}$\\
    \hline
    &C-Statistic&729.5\\
    &d.o.f.&679\\
    \enddata
    \tablecomments{Parameters of the best-fitting basic model of the
      XRISM/Resolve spectrum of Mrk 279, assuming a power law
      continuum and three Gaussian convolved \texttt{MYTorus}
      components. The models were fit to the time-averaged XRISM
      spectrum. With this model, the observed flux is $F_{2-10} = 2.1
      \times10^{-11}~\mathrm{erg~cm^{-2}~s^{-1}}$, implying a
      luminosity of $L_{\rm ion}^{\rm all} = 1.3 \times
      10^{45}~\mathrm{erg~s^{-1}}$.  \label{table:simpleparams}}
\end{deluxetable}

\begin{deluxetable*}{cccccccc} 
\tabletypesize{\footnotesize}
\label{table:physical_parameters}
\tablecaption{The Detailed Physical Model Parameters}
\tablewidth{0pt}
\tablehead{
\colhead{Parameters} &  \colhead{} & \colhead{} & \colhead{} & \colhead{}}
\startdata
    $C/\nu$ =  693/664& \texttt{powerlaw} &&\\
    $\Gamma$ & $1.67^{+0.01}_{-0.01}$ & ... & ... & ... \\
    Norm ($10^{52}$ $\mathrm{ph~s^{-1}~keV^{-1}}$) & $1.09^{+0.02}_{-0.02}$ & ... & ... & ... \\
    \hline 
    & &  \texttt{MYT$_1$} &  \texttt{MYT$_2$} &  \texttt{MYT$_3$} \\
    $r~(GM/c^2)$ & ... & $21500^{+231000}_{-11000}$ & $1560^{+5750}_{-1480}$ & $81^{+280}_{-75}$ \\
    $\theta$ & ... & $27^{+60}_{-15}$ & $34^{+53}_{-24}$ & $33^{+53}_{-9}$ \\
    Norm. ($10^7$) & ... & $10^{+3}_{-2}$ & $13^{+3}_{-3}$ & $6^{+2}_{-2}$ &\\
    $\Delta C/\Delta \nu$ & ... & -19/-3 & -16/-3 & -10/-3 \\
    $\Delta AIC$ (D.S) & ... & $-13~(3.7\sigma)$& $-10~(3.4\sigma)$ & $-4~(2.4\sigma)$ \\
    \hline 
    & &  \texttt{pion$_1$} &  \texttt{pion$_2$} &  \texttt{pion$_3$} \\
    $N_H$ ($10^{24}~\rm cm^{-2}$)  & ... & $0.056^{+0.017}_{-0.020}$ & $0.043^{+0.015}_{-0.011}$ & $0.1^-$\\
    $\log \xi$ ($\mathrm{erg~cm~s^{-1}}$) & ... & $4.40^{+0.20}_{-0.32}$ & $4.23^{+0.20}_{-0.18}$ & $4.84^{+0.14}_{-0.12}$\\
    $\sigma_v$ ($\mathrm{km~s^{-1}}$) & ... & $920^{+260}_{-590}$ & $670^{+190}_{-240}$ & $880^{+370}_{-260}$ \\
    $v_z$ ($\mathrm{km~s^{-1}}$) & ... & $-66400^{+610}_{-270}$ & $-99400^{+430}_{-170}$ & $360^{+390}_{-370}$ \\
    $f_{\rm cov}$ & ... & ... & ... & $1^{+0}_{-0.42}$ \\ 
    $\Delta C/\Delta \nu$ & ... & -12/-4 & -12/-4 & -17/-4 \\
    $\Delta AIC$ (D.S) & ... & $-3.3~(2.4\sigma)$& $-3.5~(2.4\sigma)$ & $-8.3~(3.1\sigma)$ \\
    \hline
    \hline
    $r_1$ ($GM/c^2$) Eq. \ref{eq: r_max} & ... & $1.6^{+0.5}_{-0.6} \times 10^5$ & $3.1^{+1.1}_{-0.8} \times 10^5$ & ...\\
    $r_2$ ($GM/c^2$) Eq. \ref{eq: r_min} & ... & $41^{+0.7}_{-0.3}$ & $18^{+0.2}_{-0.1}$ & ...\\
    $f_{\mathrm{v_{min}}} =  r_2 / r_1$  & ... & $2.5^{+1.5}_{-0.6} \times 10^{-4}$ & $0.6^{+0.2}_{-0.1} \times 10^{-4}$ & ... \\
    $\dot M_{\mathrm{out}}/f_\mathrm{v}$ ($M_{\odot}$/yr) Eq. \ref{eq:M_dot} & ... & $138^{+81}_{-72}$ & $300^{+175}_{-101}$ & ... \\
    $\dot E_k/f_\mathrm{v}$ $\times 10^{46}$ ($\mathrm{erg~s^{-1}}$) Eq. \ref{eq:E_k} & ... & $19^{+11}_{-10}$ & $93^{+55}_{-32}$ & ... \\
    $\dot E_k/L_{\mathrm{\rm Edd}}/f_\mathrm{v}$ & ... & $40^{+23}_{-21}$ & $195^{+110}_{-70}$ & ... \\
    $\dot E_k/L_{\mathrm{\rm Edd}}/f_\mathrm{v} \times f_{\mathrm{v_{min}}}$ & ... &  $1^{+0.6}_{-0.5} (\%)$ & $1^{+0.7}_{-0.4} (\%)$ & ... \\
    $L_{ion}^{in}$ ($\times 10^{45}$ $\mathrm{erg~s^{-1}}$) & ... & $1.27^{+0.74}_{-0.66}$ & $1.25^{+0.73}_{-0.42}$ & ... \\
    $\dot p_{\mathrm{\rm rad}}$ (dyn) Eq.\ref{eq:P_rad} & ... & $5.4^{+3.1}_{-2.8} \times 10^{34}$ & $8.3^{+4.9}_{-2.8} \times 10^{34}$ & ... \\
    $\dot p_{\mathrm{\rm out}}/f_\mathrm{v}$ (dyn) Eq.\ref{eq:P_out} & ... & $0.6^{+0.3}_{-0.3} \times 10^{38}$ & $1.9^{+1.1}_{-0.6} \times 10^{38}$ & ... \\
\enddata
\tablecomments{The upper panel (separated by the double horizontal lines) of this table presents best-fit parameters of the model, the change in C-stats/degrees of freedom ($\Delta C/\Delta \nu$), the change in AIC ($\Delta AIC$), and the detection significance (D.S) for each component. The lower panel of this table provides the key wind parameters, including launching radius ($r_1$, $r_2$), and the lower limit of volume filling factor ($f_{\mathrm{v_{min}}}$). The values of mass outflow rate $\dot M_{\mathrm{out}}/f_\mathrm{v}$, kinetic power $\dot E_{\rm k}/f_\mathrm{v}$ assuming a unity volume filling factor, the ratio of kinetic power to the Eddington limit for both unity and minimum filling factors, the input ionizing luminosity $L_{\rm ion}^{\rm in}$, radiation momentum flux $\dot P_{\mathrm{rad}}$, and outflow momentum rate assuming unity filling factor $\dot P_{\mathrm{out}}/f_\mathrm{v}$ are presented as well. The best-fit model gives X-ray flux of $F_{2-10} = 2.1 \times10^{-11}~\mathrm{erg~cm^{-2}~s^{-1}}$ in the 2-10 keV band and a total ionizing luminosity of $L_{\rm ion}^{\rm all} = 1.3 \times 10^{45}~\mathrm{erg~s^{-1}}$. Assuming $L_{\rm bol} = 2L_{\rm ion}^{\rm all}$, we estimate the bolometric luminosity of Mrk 279 to be approximately $2.6 \times 10^{45}~\mathrm{erg~s^{-1}}$, yielding an Eddington ratio of $\lambda_{\rm Edd} = L_{\rm bol}/L_{\rm Edd} = 0.5$.}
\end{deluxetable*}

\end{document}